\documentclass{article}

\PassOptionsToPackage{numbers, compress}{natbib}

 \usepackage[final]{nips_2018}

\usepackage[utf8]{inputenc} 
\usepackage[T1]{fontenc}    
\usepackage{hyperref}       
\usepackage{url}            
\usepackage{booktabs}       
\usepackage{amsfonts}       
\usepackage{nicefrac}       
\usepackage{microtype}      
\usepackage{amsmath}
\usepackage{graphicx}
\usepackage{subcaption}
\usepackage{amssymb}
\usepackage{cleveref}
\usepackage{bm}
\usepackage{color}
\usepackage{floatrow}
\newfloatcommand{capbtabbox}{table}[][\FBwidth]
\usepackage{algorithm}
\usepackage[noend]{algpseudocode}

\makeatletter
\def\BState{\State\hskip-\ALG@thistlm}
\makeatother

\title{Group induced graphical lasso allows for discovery of molecular pathways-pathways interactions}

\def\ie{\emph{i.e.}}

\DeclareMathOperator*{\minimize}{minimize}
\DeclareMathOperator{\logdet}{logdet}
\DeclareMathOperator{\tr}{tr}

\author{
Veronica Tozzo, Federico Tomasi, Margherita Squillario, Annalisa Barla \\
  Department of Informatics, Bioengineering, Robotics and System Engineering\\
  Universit\`a degli Studi di Genova \\
  Genova, Italy 16146  \\
  Corresponding author:  \texttt{veronica.tozzo@dibris.unige.it} 
}

\begin{document}

\maketitle

\begin{abstract}
  Complex systems may contain heterogeneous types of variables that interact in a
  multi-level and multi-scale manner. In this context, high-level layers may considered as
  groups of variables interacting in lower-level layers. This is particularly true
  in biology, where, for example, genes are grouped in  pathways and two types of interactions are present:
   pathway-pathway interactions and gene-gene interactions. However, from data it is only possible to measure the expression of genes while it is impossible to directly measure the activity of pathways. 
  Nevertheless, the knowledge on the inter-dependence
  between the groups and the variables allows for a multi-layer network inference,
  on both observed variables and groups, even if no direct information on the latter
  is present in the data (hence groups are considered as latent). In this paper, we
  propose an extension of the latent graphical lasso method that leverages
  on the knowledge of the inter-links between the hidden (groups) and observed
  layers. The method exploits the knowledge of group structure that influence the behaviour of observed
  variables to retrieve a two layers network. Its efficacy was tested on synthetic data to check its ability in retrieving
  the network structure compared to the ground truth. We
  present a case study on Neuroblastoma, which shows how our multi-level
  inference is relevant in real contexts to infer biologically meaningful connections.
\end{abstract}

\section{Scientific background}
The study of complex systems is typically performed through mathematical abstractions
(\emph{graphs}) that model the entities of the system as nodes and connections as edges.
A graph is a pair $\mathcal{G}=(\mathcal{E}, \mathcal{V})$
where $\mathcal{V}= \{1, \dots, d\}$ is the set of $d$ nodes and
$\mathcal{E} \subseteq \mathcal{V} \times \mathcal{V}$
is the set of edges.
In what follows we consider undirected graphs,
for which there is no distinction between $(i,j) \in \mathcal{E}$ and $(j,i)$.\\
In many applications the underlying graph is not known \textit{a priori}
but it has to be inferred from real world observations.
\emph{Gaussian graphical models} (GGMs) are a popular set of methods which aim
to solve the network inference problem \cite{lauritzen1996graphical}.
Such models assume variables to be jointly distributed according to a zero-mean
multivariate Gaussian distribution $\mathcal{N}(0, \Sigma)$
(the zero-mean assumption lets $\Sigma$ completely describe the system,
without loss of generality).
While a direct estimation of the covariance $\Sigma$ is possible,
such matrix represents the marginal distribution of the variables,
hence encoding both direct and indirect dependencies.
A different approach is to estimate its inverse $\Theta = \Sigma^{-1}$
that encodes the \emph{conditional} dependencies between variables, \ie,
$\Theta_{ij} = 0$ means that variables $i$ and $j$ are conditionally independent
given all the other variables in the graph.
Hence $\Theta$ can be considered as the weighted adjacency matrix of the graph $\mathcal{G}$.\\
A way to infer such network is through optimization of penalized maximum likelihood \cite{friedman2008sparse}
$
\logdet(\Theta) - \tr(S\Theta)~-~\alpha\|\Theta\|_{od,1}
$
where  $S = \frac{1}{n}X^\top X$ and
$\|\cdot\|_{od,1}$ is the off-diagonal $\ell_1$ norm that enforces sparsity
in the precision matrix.
Indeed, a prior on the problem is the assumption that variables interact only with few others.
Sparsity in the graph helps with the identifiability of the graph in the case
when the number of samples is much lower than the number of variables,
improves interpretability of the results and reduces noise.

Such model does not consider that real world systems
are often composed by heterogeneous entities or
some high-level mechanisms that group the entities in a structured manner~\cite{cheng2017multilevel}.
Indeed, considering such influence during the network inference may direct the analysis
towards a robust estimation of the graph.
\paragraph{Notation}
In what follows, we will denote matrices with capital letters, such as $A$.
We denote a square block of $A$ as $A_B$,
meaning that $A$ is restricted to the rows and columns corresponding to indices in the set $B$.
Similarly, we denote a non-square block as $A_{BC}$,
considering rows in the set $B$ and columns in the set $C$.
$A_{ij}$ indicates the entry of $A$ corresponding to the $i$-th row and $j$-th column. 

\section{Group induced graphical lasso }
We want to infer a two-layers networks in which one layer model connections between observed variables and the other the connections between groups of these observed variables. Note that the activity of the groups cannot be directly measured and it is, therefore, \emph{latent}. Nevertheless, we assume that the membership of each observed variable to one or more groups is known \textit{a priori}.
The goal of a advanced analysis is to provide a more complete overview of the system even though  we do not expect the connections between groups to exhaustively explain the connections between observed variables. Hence we allow connections between variables that belong to different groups.
To this aim we extend a state-of-the-art method for latent graphical inference (LGL) \cite{chandrasekaran2010latent,yuan2012} that takes into account the influence of latent factors without any assumption on their identities.
The proposed method is Group Induced Graphical Lasso (GIGL) that considers structured regularisation over the network links to consider the group memberships of the variables.
\subsection{The model} 
Consider $N$ observations in $O$ variables represented as a matrix
$X \in \mathbb{R}^{N \times O}$ drawn from a multivariate Gaussian distribution
$\mathcal{N}(0, \tilde{\Sigma}_O)$.
Observed variables $O$ describe the available samples,
but their interaction is perturbed by the influence of higher-level interactions of $H$ groups which activity is unobserved (\emph{hidden}).
The matrix perturbation is 
$\tilde{\Sigma}_O = \Sigma_O - \Sigma_{OH}\Sigma_H^{-1}\Sigma_{OH}^\top$
where $\Sigma_O$ is the true underlying covariance of the observed variables.
We aim at inferring the global precision matrix of the system given
the observations $X$ and the memberships of the $O$ variables in $H$ groups.
These memberships are encoded into a binary matrix $G \in \{0,1\}^{o \times h}$
where $G_{oh} = 1$ if the observed variable $o$ belongs to the group $h$ and $0$ otherwise.
Note that groups can overlap.
Our goal is to estimate the precision matrix $\Theta$ of the following form:
\small
\begin{equation*}
\label{matrixform}
\Theta = \Sigma^{-1} = \left[
\begin{array}{c|ccc}
& \hspace{-.5em} &                  & \hspace{-.5em} \\[-.6em]
\Theta_{H}  & \hspace{-.5em} & \Theta_{OH}^\top & \hspace{-.5em} \\[.5em]
\hline
& \hspace{-.5em} &                  & \hspace{-.5em} \\
\Theta_{OH} & \hspace{-.5em} & \Theta_{O}       & \hspace{-.5em} \\
& \hspace{-.5em} &                  & \hspace{-.5em} \\
\end{array}\right].
\end{equation*}
\normalsize
where $\Theta_H$ and $\Theta_O$ represent the precision matrices between groups and variables respectively and $\Theta_{OH}$ is a possibly non-squared matrix.
Note that a non-zero entry in the precision matrix $\Theta_{OH}$ should be found in correspondence of non-zero entries of $G$. 
Moreover, the model does not prevent the inference new connections between variables and groups, since there may be some dependency which is not known a priori.

The inference problems translates into the optimisation of a non-convex function
$\mathcal{F}(\Theta, \widetilde{S}_0, G)$
\small
\begin{align}\label{eq:main problem}
\begin{split}
\underset{\begin{subarray}{c}
	\Theta \in \mathbb{R}^{(O+H)\times(O+H)}\\
	\Theta \succ 0
	\end{subarray}} {\minimize} &~~\mathcal{F}(\Theta, \tilde(S)_0, G) \\
=  \underset{\begin{subarray}{c}
	\Theta \in \mathbb{R}^{(O+H)\times(O+H)}\\
	\Theta \succ 0
	\end{subarray}} {\minimize} &- \logdet(\Theta) + \tr(S\Theta) + \lambda\|\Theta_O\|_{1, od} +  + \eta\|\Theta_H\|_{1, od} +  \mu \bar{G} \|\Theta_{OH}\|_1
\end{split}
\end{align}
\normalsize
where $S \in \mathbb{R}^{O+H}$, $S_O = \frac{1}{N}X^\top X$ and $\bar{G} = 1 - G$.

Problem~\eqref{eq:main problem} depends on three hyper-parameters $\lambda$, $\eta$ and $\mu$,
which regulate the sparsity of the blocks of the matrix $\Theta$.
The sparsity on the block $\Theta_{OH}$ is also enriched by the structured regularisation given by $\bar{G}$. 
\Cref{sec:EM} includes the optimisation algorithm and related mathematical derivations.

\section{Experiments}
We tested the effectiveness of GIGL on both synthetic and real data.
Synthetic experiments show its ability in retrieving the network structure. Results on real data show how GIGL, through the imposition of prior knowledge,
allows some latent patterns to emerge.
We performed a model selection procedure to choose the best hyper-parameters
$\lambda$, $\mu$ and $\eta$ based on the data at hand.
The procedure was performed on a refined grid using the average log-likelihood score
of the model with respect to test data on a 3-fold cross validation.
The log-likelihood is computed as:
\begin{equation}
\logdet(\Theta_{obs}) - \tr(\tilde{S}_{ts}\Theta_{obs}),
\end{equation}
where $S_{ts}$ is the empirical covariance matrix of the test data and
$\Theta_{obs} = \Theta_O - \Theta_{OH}\Theta_{H}^{-1}\Theta_{OH}^\top$.

\begin{figure}
	\begin{subfigure}{0.24\textwidth}
		\includegraphics[width=1\textwidth]{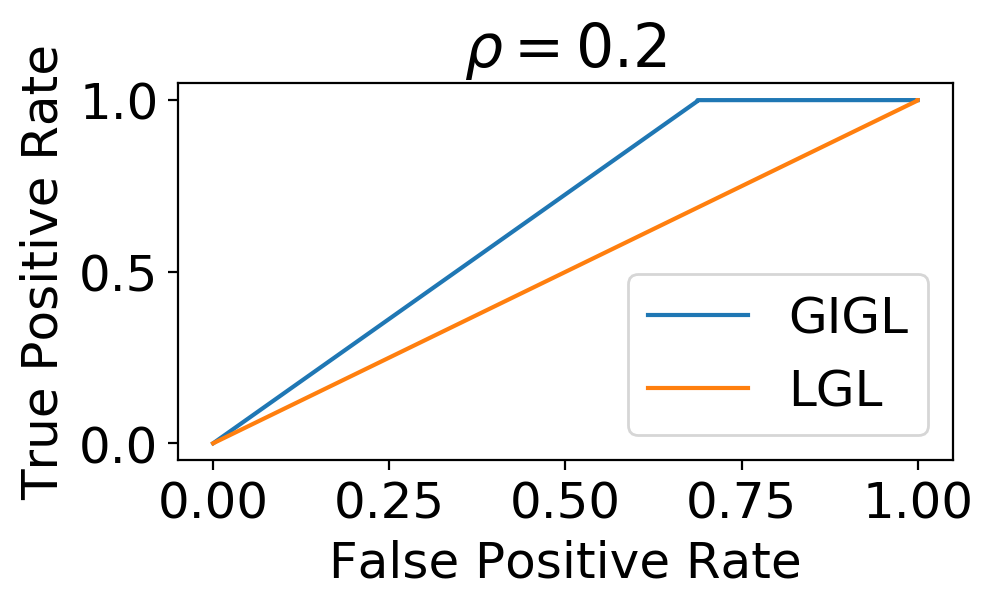}
	\end{subfigure}
	\begin{subfigure}{0.24\textwidth}
		\includegraphics[width=1\textwidth]{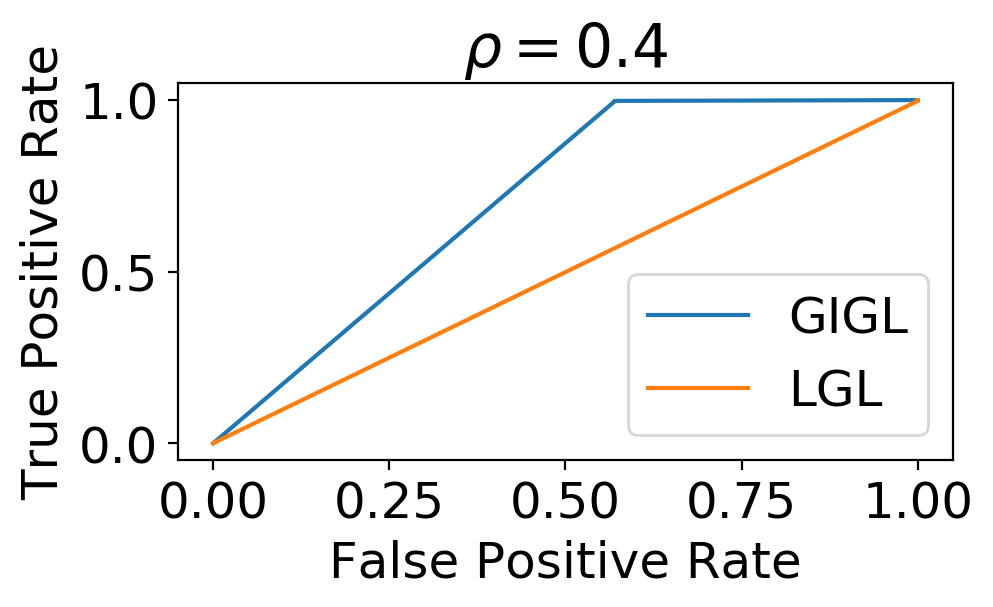}
	\end{subfigure}
	\begin{subfigure}{0.24\textwidth}
		\includegraphics[width=1\textwidth]{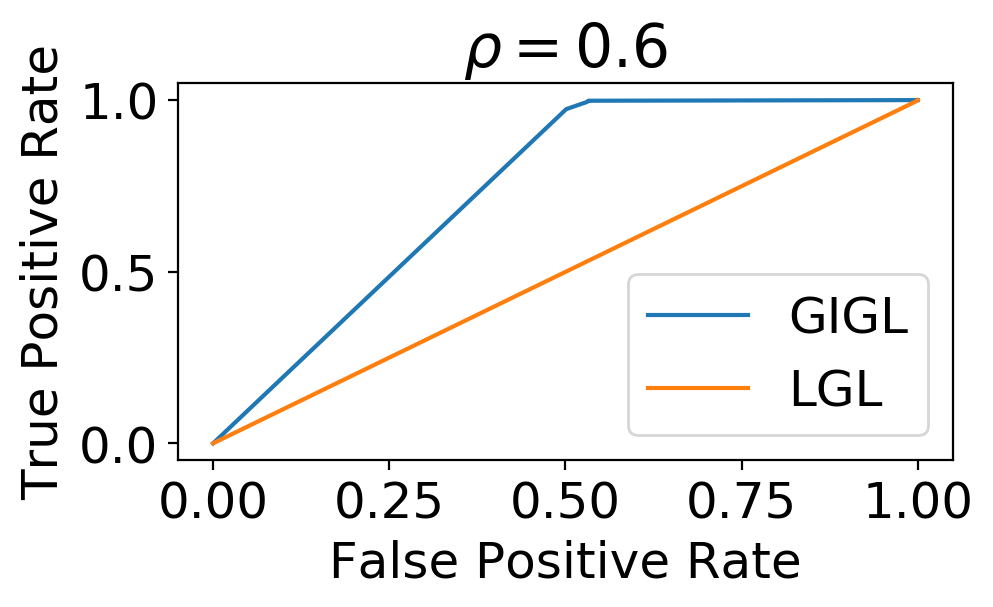}
	\end{subfigure}
	\begin{subfigure}{0.24\textwidth}
		\includegraphics[width=1\textwidth]{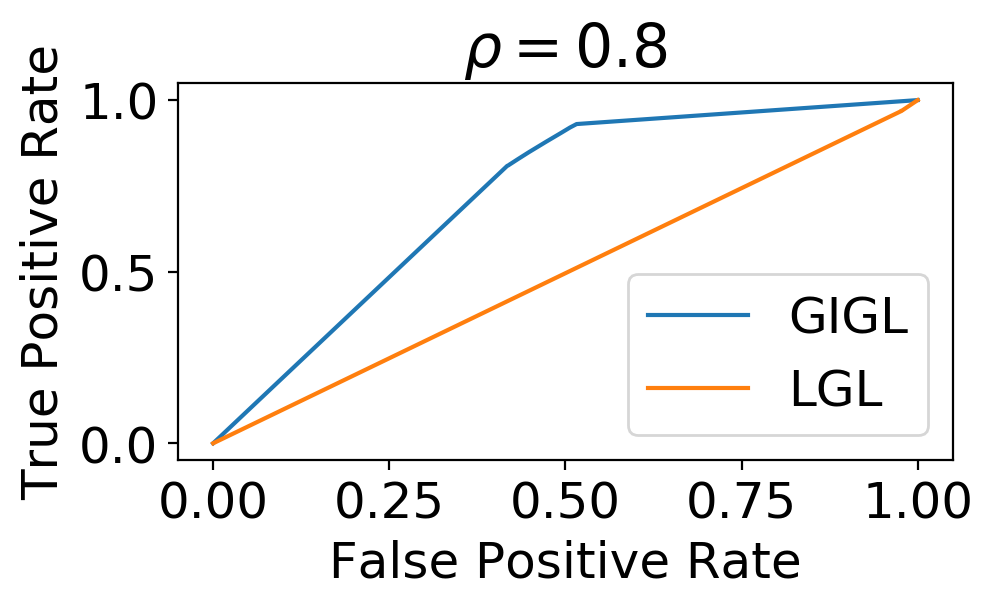}
	\end{subfigure}
\begin{tabular}{cccc}
	$AUC_{\rho=0.2} = 0.66$\hspace{0.3cm} & $AUC_{\rho=0.4} = 0.71$\hspace{0.3cm} & $AUC_{\rho=0.6} = 0.74$ \hspace{0.3cm}& \hspace{0.5cm}$AUC_{\rho=0.8} = 0.72$
\end{tabular}
\caption{ROC curves of GIGL and LGL methods. The experiments were performed with an increasing sparsity of the ground truth network. The AUC of GIGL increases as the sparsity increases, while for LGL it remains stable at $0.5$ denoting that LGL tends to be less sparse than GIGL.}
\label{fig:res_synth}
\end{figure}

\paragraph{Synthetic experiments}
We generated four different data sets with increasing sparsity $\rho$ of the connections, with $\rho \in \{0.2, 0.4, 0.6, 0.8\}$. Each data set has 100 latent variables and 200 observed variables. 
%
The blocks of $\Theta$ are generated such that both the global matrix
and the difference $\tilde{\Theta} = \Theta_O - \Theta_{OH}\Theta_{H}^{-1}\Theta_{OH}^\top$ are positive definite.
We drew 500 samples from a multivariate Gaussian distribution
$\mathcal{N}(0, \tilde{\Theta}^{-1})$ 
to estimate the empirical covariance matrix, which is the input of GIGL.
Also, we supplied to GIGL the mask for the non-zero positions of $\Theta_{OH}$.
The performance of GIGL and LGL were evaluated in terms of structure recovery,
based on the Receiving Operating Characteristic (ROC) curve and Area Under the Curve (AUC) (\Cref{fig:res_synth}). 
Note that the ROC curve requires probabilities/confidence interval to be computed. 
To assess the robustness of the results we estimated a confidence interval by repeating the experiments 10 times.
We then computed the average of all binarised resulting matrices $\Theta_i$ for $i=1, \dots 10$.  We interpret the average matrix $\bar{\Theta}$ as the probability of each edge to exist. 

\begin{figure}[h]
	\begin{floatrow}
		\ffigbox{%
			\includegraphics[width=0.4\textwidth]{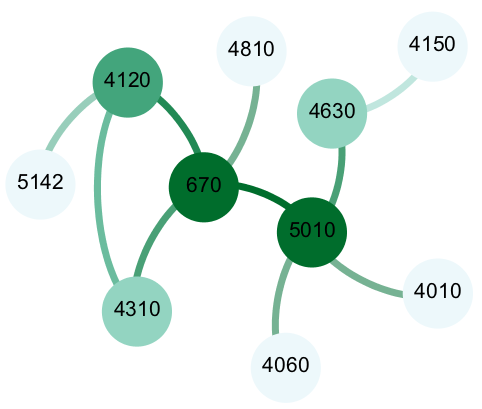}
		}{%
			\caption{KEGG pathways connections on Neuroblastoma estimated by GIGL.\@
			The darker colour of the node denotes its degree while the darker colour of the edge denotes the probability of its existence.}
			\label{fig:pathways_network}
		}
		\capbtabbox{%
			\footnotesize
			\begin{tabular}{rl}
				\textbf{No.} & \textbf{Description}\\
				\hline
				4060 & Cytokine cytokine receptor interaction\\
				5010 & Alzheimer's disease \\
				670 & One carbon pool by folate \\
				4630 &  jak STAT signaling pathway \\
				4120  & Ubiquitin mediated proteolysis \\
				5142 & Chagas disease american trypanosomiasis\\
				4010 & MAPK signaling pathway \\
				4150 & mTor signaling pathway \\
				4310 & wnt signaling pathway \\
				4810 & Regulation of actin cytoskeleton \\
				\hline
			\end{tabular}
		}{%
			\caption{KEGG pathways description of the network nodes in \Cref{fig:pathways_network}}%
		}
	\end{floatrow}
\end{figure}

\paragraph{Neuroblastoma: a case study}
We applied our method on a TCGA neuroblastoma RNA-Seq dataset\footnote{Available at https://portal.gdc.cancer.gov/projects/TARGET-NBL.}.
For computational ease, we considered a subset of genes known in literature
to be involved in Neuroblastoma disease based on Phenopedia~\cite{yu2010phenopedia}.
The resulting list of 203 genes was provided to Webgestalt~\cite{Wang2013WEB}
for a functional characterisation through a gene enrichment analysis.
We ended up considereing 116 KEGG (Kyoto Encyclopedia of Genes and Genomes) \cite{kanehisa2000kegg} pathways where the subset of genes was found enriched.\\
We then applied GIGL and LGL to this data by imputing the empirical covariance matrix of the gene expression and the membership of each gene to one or more pathways.
Given the non-convexity of our model 
we optimise the model 20 times with different initialisations,
which may lead to different solutions.
Finally, we retain the links present at least 70\% of the 20 times.
\Cref{fig:pathways_network} shows the pathways-pathways interactions.
The gene networks for both GIGL and LGL are included in \Cref{sec:figures},
due to space constraints. 

\section{Discussion}
Synthetic experiments show that our method is able to correctly identify the links between latent variables acting in the system also with an increasing sparsity.
As shown in \Cref{fig:res_synth}, LGL always finds dense blocks in the latent variables, while GIGL better identify the links.
We also note how an increase in sparsity results in a more accurate identification
of missing edges.
Considering the application on neuroblastoma,
we retrieved two gene interactions network,
one for both GIGL and LGL (in Appendix~\ref{sec:figures} \Cref{fig:gene_network,fig:gene_network_yuan}).
GIGL is able to identify pathways-pathways interactions (\Cref{fig:pathways_network}).
The difference between the sparsity level of the networks as estimated by the two methods is also more generally visible in
\Cref{fig:adjacency_ours,fig:adjacency_yuan}. 
The inferred co-expression networks
(\Cref{fig:gene_network,fig:gene_network_yuan})
include four common genes that emerge above others.
These genes are PLEKHA4, IL6, S100B and NTRK. While the relevance of IL6 \cite{totaro2013impact, zhao2018serum} and NTRK \cite{lipska2009c} in neuroblastoma is a known fact, the role of the remaining two genes is still under investigation. 
Differently from PLEKHA4 which is poorly annotated, S100B is a well characterized gene whose chromosomal rearrangements and altered expression are known to be implicated in several neurological, neoplastic, and other types of diseases, including Alzheimer's disease, Down's syndrome and epilepsy.
In GIGL network (\Cref{fig:gene_network}) there are more hub genes (\ie, genes highly connected) than in LGL (\Cref{fig:gene_network_yuan}).
The hubs genes are: MICA, EGFR, NEFL, MYO10, VDR, HFE, HLA-C, GHR, PDLIM1, APOE, DAB2, PALU, CEBPA, IL-33. The involvement in Neuroblastoma of the first 7 genes of this list is present in literature \cite{borriello2016more, capasso2014common, wang2018mir, bini2012new, mrowczynski2017hfe, cheng1996lack}.
Also \Cref{fig:pathways_network} shows that ``Alzheimer’s disease'' and ``One carbon pool by folate'' are the two most strongly connected pathways.
Folic acid has been recently connected to childhood cancer \cite{moulik2017folic}, 
while the one-carbon pathway was linked to Alzheimer's \cite{fuso2011dna}.
\Cref{fig:pathways_network} also shows a clique between  ``Wnt signaling'',  ``Ubiquitin mediated proteolysis'' and ``One carbon pool by folate'' pathway.
It is known that WNT signaling pathway plays significant roles in the survival, proliferation, and differentiation of human neuroblastoma \cite{suebsoonthron2017inhibition} and ``Ubiquitin mediated proteolysis'' is crucial in the regulated degradation of proteins involved in neuroblastoma proliferation and survival~\cite{hammerle2013targeting}.\\
In this paper we proposed an extension of the latent graphical lasso that considers prior knowledge on the interlinks to retrieve connections between both observed and latent entities. Our method can be used to look at different network topologies that can emerge by imposing different group priors on the observed variables.
The method allows for a straightforward application to real-world data,
not limited to biological data sets. 
Indeed, while propose it for groups of variables this may also be used for general known connections between latent and observed variables that do not translate into memberships. 
In this work we applied our method only on a small subset of Neuroblastoma data, we aim at applying to a larger dataset and to compare the priors between different biological groups. We also plan to compare it with other methods that perform multi-layer inference \cite{cheng2017multilevel, lin2016penalized} that have different assumptions on data generation to look for evidence that emerges with more than one methodology. 

\bibliographystyle{plain}
\bibliography{bibliografia.bib}

\newpage
\appendix
\section{Optimization algorithm}\label{sec:EM}
Given problem~\eqref{eq:main problem}, we report the mathematical steps required to reach the form of \Cref{alg:optimization}.
In particular the problem has $H$ latent variables that need to be estimated from input data.
This purpose is reached through Expectation Maximization. 

\paragraph{Expectation step}
We compute the expected value of the penalized negative log-likelihood function given $S_O$ and $K^{t-1}$. 
\small\begin{align}
\begin{split}
\mathcal{Q}(K | K^{t-1}) &= \mathbb{E}_{X_H|X_O, \Theta^{t-1}}\bigg[- \text{log det}(\Theta) + \text{tr}(S\Theta) + \lambda\|\Theta_O\|_{1, od} +  + \eta\|\Theta_H\|_{1, od} +  \mu \bar{G} \|\Theta_{OH}\|_1\bigg] \\
& = - \text{log det}(\Theta) + \text{tr}\big(\mathbb{E}_{X_H|X_O, \Theta^{t-1}}(S) \Theta\big) + \lambda\|\Theta_O\|_{1, od} +  + \eta\|\Theta_H\|_{1, od} +  \mu \bar{G} \|\Theta_{OH}\|_1
\end{split}
\end{align}\normalsize

Let $\Sigma^t = (\Theta^{t-1})^{-1}$. Then, 
\small
\begin{equation}
\mathbb{E}_{X_H| X_O, \Theta^{t-1}}(S_{OH}) = \widetilde{S}_{OH} = S_O(\Sigma^t_O)^{-1}\Sigma^t_{OH}
\end{equation}
\begin{equation}
\mathbb{E}_{X_H|X_O, \Theta^{t-1}}(S_H) = \widetilde{S}_{H} = \Sigma^t_O - (\Sigma^t_{OH})^\top(\Sigma^t_O)^{-1}\Sigma^t_{OH} + (\Sigma^t_{OH})^\top(\Sigma^t_O)^{-1}S_O(\Sigma^t_O)^{-1}\Sigma^t_{OH}
\end{equation}\normalsize

\paragraph{Maximization step}
After we compute the expectation we can reconstruct the global matrix $S$ where we need to rectify $S_O$ by adding the influence of the hidden factors.
Hence,
\small \begin{equation}
\widetilde{S}_{O} =  S_O + \widetilde{S}_{OH}(\widetilde{S}_{H})^{-1}\widetilde{S}_{OH}^\top
\end{equation} \normalsize
ending with
\small
\begin{equation}
\label{matrixform}
\tilde{S} = \left[\begin{array}{c|ccc}
&\hspace{-.5em}&&\hspace{-.5em}\\[-.6em]
\widetilde{S}_{H} &\hspace{-.5em}&\widetilde{S}_{OH}^\top&\hspace{-.5em}\\[.5em]
\hline
&\hspace{-.5em}&&\hspace{-.5em}\\
\widetilde{S}_{OH}&\hspace{-.5em}&\widetilde{S}_{O}&\hspace{-.5em}\\
&\hspace{-.5em}&&\hspace{-.5em}\\
\end{array}\right].
\end{equation}
\normalsize
The maximization step translates into a weighted graphical lasso~\cite{friedman2008sparse} that can be easily be solved as the standard graphical lasso by using a matrix regulariser. This matrix regulariser will enforce the group structure and will allow pattern to emerge.  

\small
\begin{align}
\begin{split}
\Theta^t = \underset{\begin{subarray}{c}
	\Theta \in \mathbb{R}^{(O+H)\times(O+H)}\\
	\Theta \succ 0
	\end{subarray}} {\text{minimize}} - \text{log det}(\Theta) + \text{tr}(\tilde{S}\Theta) + \lambda\|\Theta_O\|_{1, od} +  + \eta\|\Theta_H\|_{1, od} +  \mu \bar{G} \|\Theta_{OH}\|_1
\end{split}
\end{align}
\normalsize
\small
\begin{algorithm}
	\caption{GIGL($\tilde{S}_O, G, \lambda, \eta, \mu)$}\label{alg:optimization}
	\begin{algorithmic}[1]
		\small
		\State $\bar{G} = 1 - G$
		\For {$t = 1, \dots, \text{max\_iter}$}
		\State $\Sigma^t = (\Theta^{t-1})^{-1}$
		\State $\widetilde{S}_{OH} = S_O(\Sigma^t_O)^{-1}\Sigma^t_{OH}$
		\State $\widetilde{S}_{H} = \Sigma^t_O - (\Sigma^t_{OH})^\top(\Sigma^t_O)^{-1}\Sigma^t_{OH} + (\Sigma^t_{OH})^\top(\Sigma^t_O)^{-1}S_O(\Sigma^t_O)^{-1}\Sigma^t_{OH}$
		\State $\widetilde{S}_{O} =  S_O + \widetilde{S}_{OH}(\widetilde{S}_{H})^{-1}\widetilde{S}_{OH}^\top$
		\State $
		\label{matrixform}
		\tilde{S} = \left[\begin{array}{c|ccc}
		&\hspace{-.5em}&&\hspace{-.5em}\\[-.6em]
		\widetilde{S}_{H} &\hspace{-.5em}&\widetilde{S}_{OH}^\top&\hspace{-.5em}\\[.5em]
		\hline
		&\hspace{-.5em}&&\hspace{-.5em}\\
		\widetilde{S}_{OH}&\hspace{-.5em}&\widetilde{S}_{O}&\hspace{-.5em}\\
		&\hspace{-.5em}&&\hspace{-.5em}\\
		\end{array}\right].
		$
		\State $R = ones(O+H, O+H)$
		\State $R = R - diag(R)$
		\State $R_O = \lambda*R_O, R_{OH} = \mu*\bar{G}, R_H = \eta*R_H$
		\State  $\Theta^t = \underset{\begin{subarray}{c}
			\Theta 
			\end{subarray}} {\text{minimize}} - \text{log det}(\Theta) + \text{tr}(\tilde{S}\Theta) + R\|\Theta\|_{1}$
		\EndFor
		\Return $\Theta$
	\end{algorithmic}
\end{algorithm}
\normalsize

\clearpage
\section{Retrieved Neuroblastoma networks }\label{sec:figures}
\begin{figure}[htp]
	\begin{subfigure}{0.45\textwidth}
	\centering
	\includegraphics[width=1\textwidth]{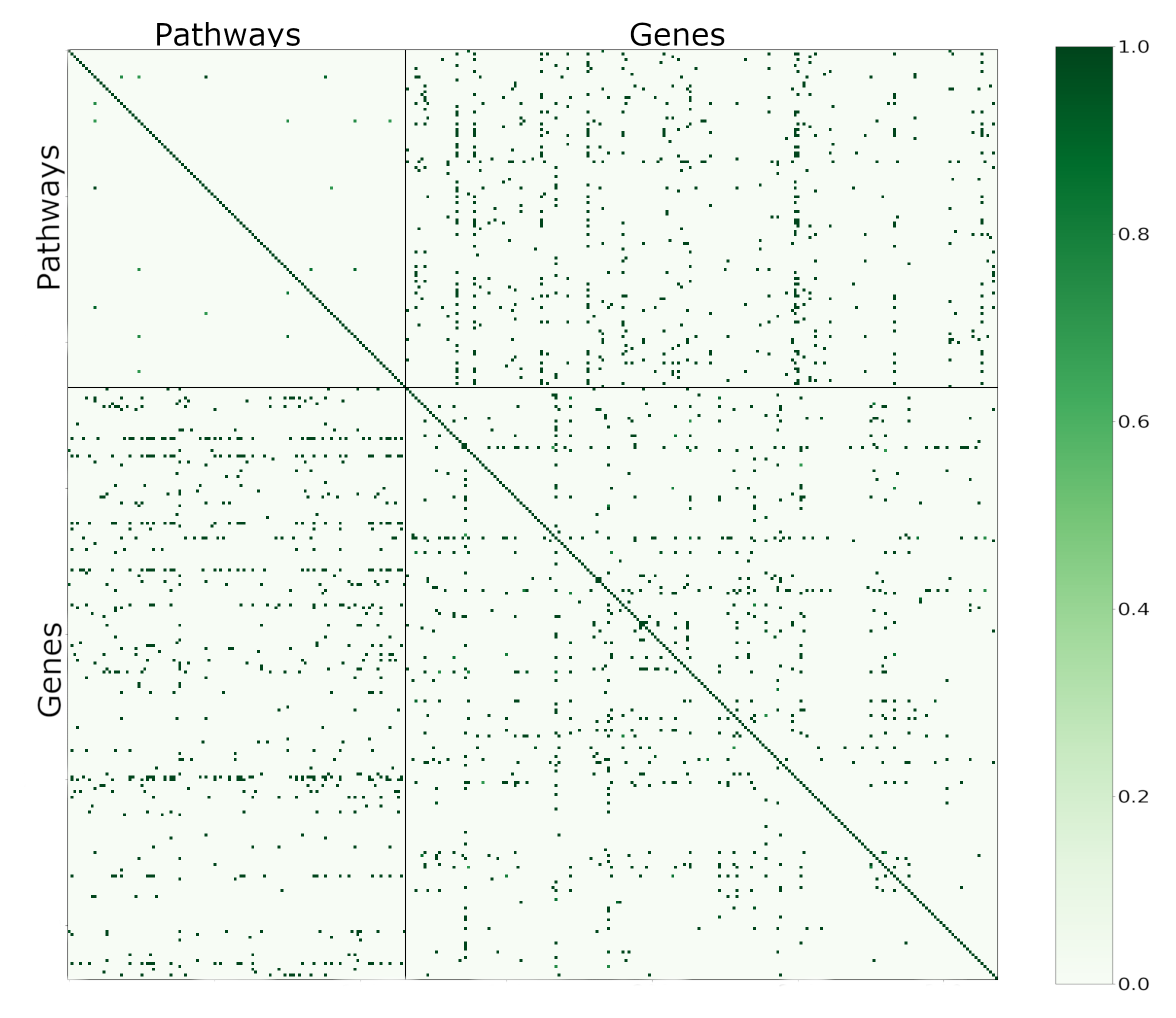}
	\caption{GIGL adjacency matrix of neuroblastoma gene co-expression and pathways-pathways interactions network. }
	\label{fig:adjacency_ours}
	\end{subfigure}
\qquad
\begin{subfigure}{0.45\textwidth}
	\centering
	\includegraphics[width=1\textwidth]{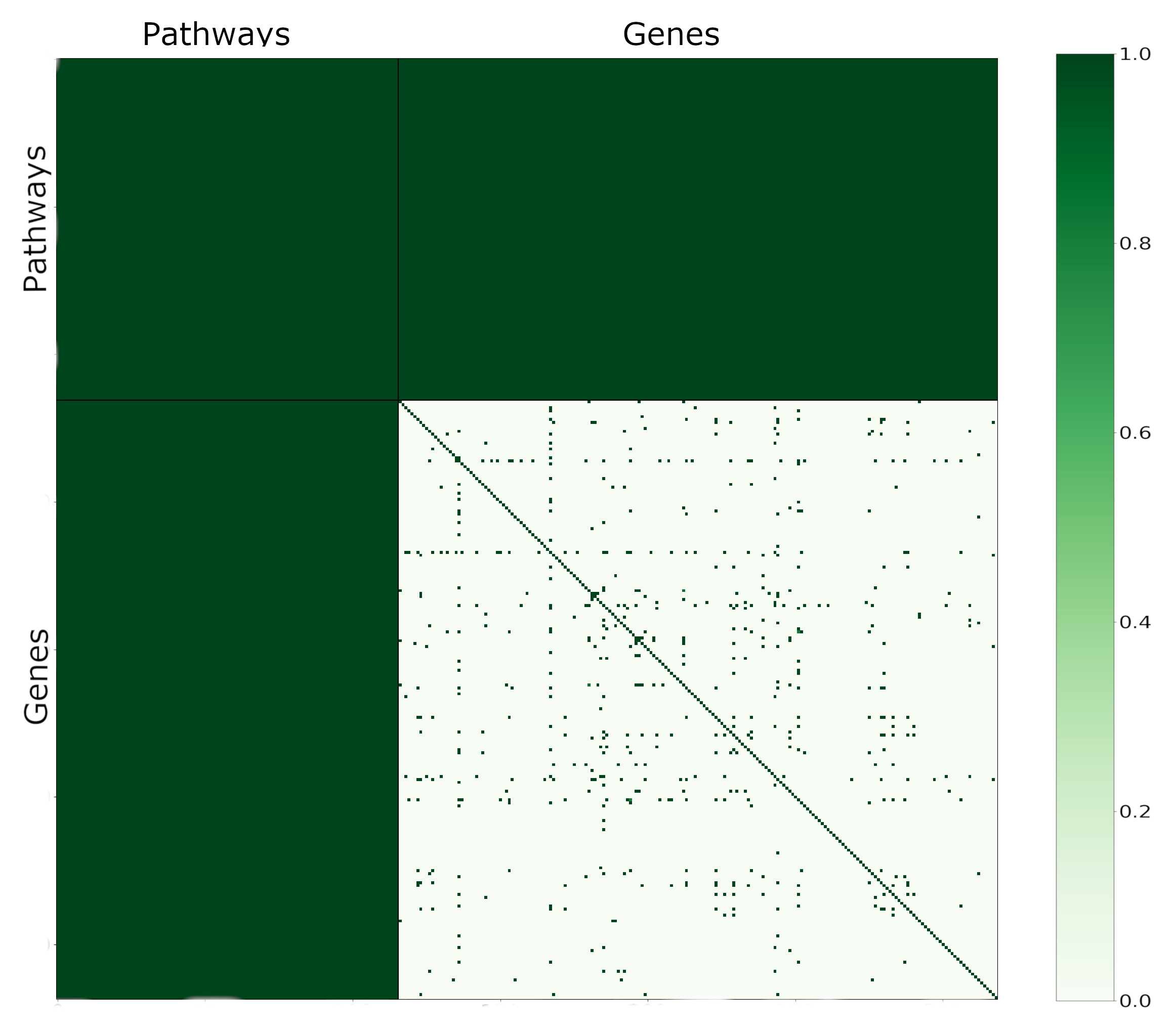}
	\caption{LGL adjacency matrix of neuroblastoma gene co-expression. LGL method does not enforce sparsity in the latent sub-block of the matrices, hence the pathways-pathways interaction as well as the pathways-genes interactions are completely dense.}
	\label{fig:adjacency_yuan}
\end{subfigure}
\caption{Comparison between the adjacency matrix obtained on Neuroblastoma data using GIGL and LGL model.}
\end{figure}

We applied both GIGL and LGL to Neuroblastoma data to show how our method
allows to extract information on group links rather than only on gene co-expression networks.
This can be seen in \Cref{fig:adjacency_ours,fig:adjacency_yuan}.  Here we show the adjacency matrices that are thresholded at 70\% of occurrences of the edges. It is easy to see that LGL (\Cref{fig:adjacency_yuan}) retrieves a completely dense network on the pathways interactions while GIGL (\Cref{fig:adjacency_ours}) is able to retrieve some strong connections between them. 

Interestingly, the cross-validated hyper-parameter $\eta$
is so high to enforce the Pathways-Genes interaction sub-matrices to contain only the prior knowledge links. This proves that this group structure is present in the data and the model automatically recognises it using a high regularisation. 

\begin{figure}[h]
	\includegraphics[width=1\textwidth]{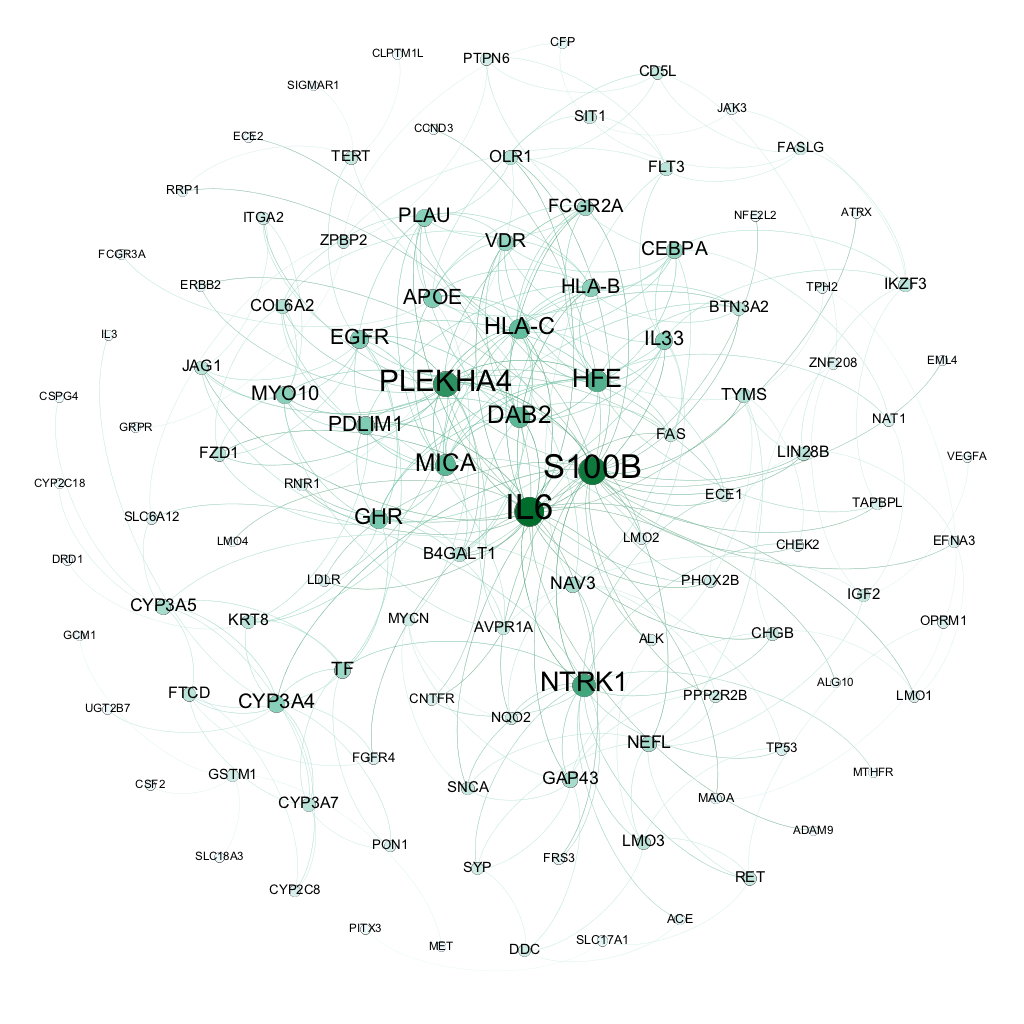}
	\caption{Neuroblastoma gene network estimated by GIGL. The more central and bigger nodes are the ones with highest degree.}
	\label{fig:gene_network}
\end{figure}
\begin{figure}[h]
	\includegraphics[width=1\textwidth]{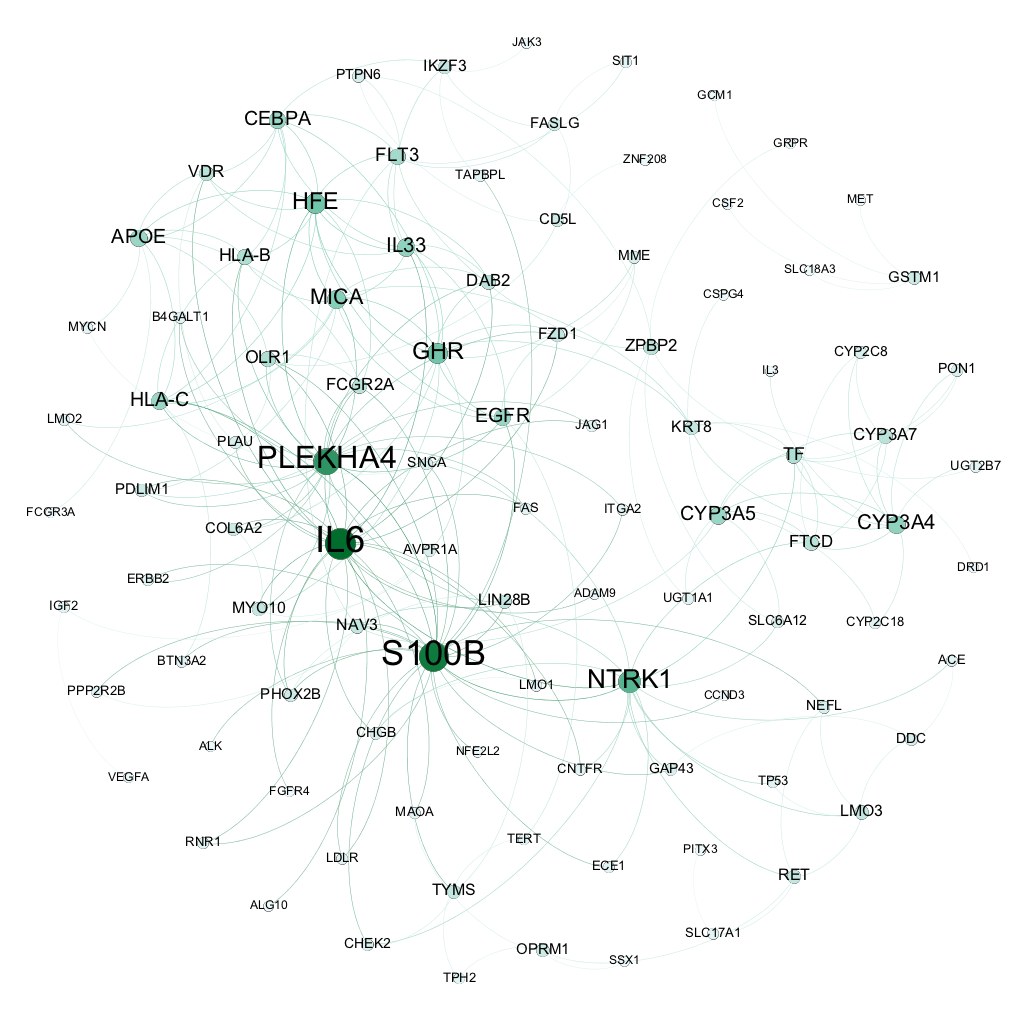}
	\caption{Neuroblastoma gene network estimated by LGL \cite{yuan2012}.
	The more central and bigger nodes are the ones with highest degree. }
	\label{fig:gene_network_yuan}
\end{figure}

\Cref{fig:gene_network,fig:gene_network_yuan} depict the gene co-expression networks obtained with GIGL and LGL.
While there exists some shared hubs across the two networks,
GIGL estimates more hubs emerging as relevant which were not found by LGL.
It seems that the grouping of pathways enforces these genes to emerge. Interestingly, the majority of such genes can be found in literature to have  well-known connections with Neuroblastoma.
\end{document}